*Ivan BORODII*

*Postgraduate Student at the Department of Computer Systems and Networks, Ternopil Ivan Puluj National Technical University, 56, Ruska Str., Ternopil, 46001, Ukraine, ivanborodii@tntu.edu.ua*
*ORCID: 0009-0005-4986-4429*
*Scopus Author ID: 60153325800*

*Halyna OSUKHIVSKA*

*Candidate of Technical Sciences, Associate Professor, Head of the Department of Computer Systems and Networks, Ternopil Ivan Puluj National Technical University, 56, Ruska Str., Ternopil, UA46001, Ukraine, osukhivska@tntu.edu.ua*
*ORCID: 0000-0003-0132-1378*
*Scopus Author ID: 57373480200*

# RESEARCH ON THE EFFICIENCY OF DATA LOADING AND STORAGE IN DATA LAKEHOUSE ARCHITECTURES FOR THE FORMATION OF ANALYTICAL DATA SYSTEMS

*The paper presents a study of the efficiency of loading and storing data in the three most common Data Lakehouse systems, including Apache Hudi, Apache Iceberg, and Delta Lake, using Apache Spark as a distributed data processing platform. The study analyzes the behavior of each system when processing structured (CSV) and semi-structured (JSON) data of different sizes, including loading files up to 7 GB in size.*

***The purpose of the work*** *is to determine the most optimal Data Lakehouse architecture based on the type and volume of data sources, data loading performance using Apache Spark, and disk size of data for forming analytical data systems.*

***The methodology*** *of the research covers the development of four sequential ETL processes, which include reading, transforming, and loading data into tables in each of the Data Lakehouse systems. The efficiency of each Lakehouse was evaluated according to two key criteria: data loading time and the volume of tables formed in the file system.*

***The scientific novelty.*** *For the first time, a comparison of performance and data storage in Apache Iceberg, Apache Hudi, and Delta Lake Data Lakehouse systems was conducted to select the most relevant architecture for building analytical data systems. The practical value of the study consists in the fact that it assists data engineers and architects in choosing the most appropriate Lakehouse architecture, understanding the balance between loading performance and storage efficiency.*

***Conclusions.*** *Experimental results showed that Delta Lake is the most optimal architecture for systems where the priority is the speed of loading data of any volume, while Apache Iceberg is most appropriate for systems where stability and disk space savings are critical. Apache Hudi proved ineffective in data loading and storage evaluation tasks but could potentially be effective in incremental update and streaming processing scenarios.*

***Key words:*** *Apache Hudi, Apache Iceberg, Apache Spark, Data Lakehouse, Delta Lake, data processing, data loading.*

**Іван БОРОДІЙ**
аспірант кафедри комп'ютерних систем та мереж, Тернопільський національний технічний університет імені Івана Пулюя, вул. Руська, 56, м. Тернопіль, Україна, 46001
*ORCID:* 0009-0005-4986-4429
*Scopus Author ID:* 60153325800

**Галина ОСУХІВСЬКА**
кандидат технічних наук, доцент, завідувач кафедри комп'ютерних систем та мереж, Тернопільський національний технічний університет імені Івана Пулюя, вул. Руська, 56, м. Тернопіль, Україна, 46001
*ORCID:* 0000-0003-0132-1378
*Scopus Author ID:* 57373480200

# ДОСЛІДЖЕННЯ ЕФЕКТИВНОСТІ ЗАВАНТАЖЕННЯ ТА ЗБЕРІГАННЯ ДАНИХ У АРХІТЕКТУРАХ DATA LAKEHOUSE ДЛЯ ФОРМУВАННЯ АНАЛІТИЧНИХ СИСТЕМ ДАНИХ

*У статті представлено дослідження ефективності завантаження та збереження даних у трьох найпоширеніших системах Data Lakehouse, серед яких Apache Hudi, Apache Iceberg та Delta Lake, використовуючи Apache Spark як платформу розподіленої обробки даних. У межах дослідження проаналізовано поведінку кожної системи під час обробки структурованих (CSV) та напівструктурованих (JSON) даних різного обсягу, включаючи завантаження файлу обсягом до 7 ГБ.*

***Метою роботи*** *є визначення найоптимальнішої Data Lakehouse архітектури на основі типу і обсягу джерел даних, продуктивності завантаження даних засобом Apache Spark, дискового розміру даних для формування аналітичних систем даних.*

***Методологія*** *дослідження охоплює побудову чотирьох послідовних ETL процесів, які включають зчитування, трансформацію та завантаження даних у таблиці кожної з Data Lakehouse систем. Ефективність кожної з Lakehouse оцінювалась за двома ключовими критеріями: час завантаження даних та обсяг сформованих таблиць у файловій системі.*

***Наукова новизна.*** *Вперше здійснено порівняння продуктивності та збереження даних в Data Lakehouse системах Apache Iceberg, Apache Hudi та Delta Lake з метою вибору найбільш релевантної архітектури для формування аналітичних систем даних. Практична цінність дослідження полягає в тому, що воно допомагає інженерам і архітекторам даних вибрати найбільш відповідну Lakehouse архітектуру, розуміючи баланс між продуктивністю завантаження та ефективністю зберігання.*

***Висновки.*** *Експериментальні результати продемонстрували, що Delta Lake є найбільш оптимальною архітектурою для систем, де пріоритетом є швидкість завантаження даних будь-якого об'єму, тоді як застосування Apache Iceberg є найбільш доцільним у системах, де критичними є стабільність та економія дискового простору. Apache Hudi виявився неефективним у задачах оцінки завантаження та збереження даних, але потенційно може бути ефективним у сценаріях інкрементальних оновлень та потокової обробки.*

***Ключові слова:*** *Apache Hudi, Apache Iceberg, Apache Spark, Data Lakehouse, Delta Lake, обробка даних, завантаження даних.*

**Relevance of the Problem.** In modern digital world, the amount of data is growing rapidly, creating new challenges for information storage and processing systems. Advanced software systems such as the Internet of Things (IoT), social networks, telemetry, and monitoring systems generate large amounts of data in real time, requiring reliable and scalable technologies for storing, accessing, and transforming information. Analytical data systems also depend on the ability to efficiently load, store, and process datasets of different volumes and formats to support reporting, forecasting, and machine learning tasks, which further increases the demands on modern data architectures. As a result, traditional data warehouses are gradually being replaced by more flexible and productive solutions, particularly the Data Lakehouse concept, which combines the advantages of Data Lakes and Data Warehouses and serves as the technological foundation for scalable and reliable analytical operations (Armbrust, 2021).

The Data Lakehouse architecture provides the ability to simultaneously store structured and semi-structured data, support transactions, manage metadata, and efficiently execute analytical queries. In this context, table storage formats such as Apache Hudi, Apache Iceberg, and Delta Lake deserve special attention, as they are the most common solutions in the implementation of Lakehouse systems (Jain, 2023).

Apache Iceberg is a Netflix product that is an open-source tabular format designed to work with large volumes of analytical data in Lakehouse-type repositories. The Iceberg architecture separates physical storage and metadata, enabling efficient table version management, change history support, and dynamic data partitioning (Apache Iceberg, 2025).

Developed by Uber, Apache Hudi is a data storage system designed to handle large volumes of constantly changing information. It supports two file processing modes: Copy-on-Write for fast reading and Merge-on-Read for efficient updates (Apache Hudi, 2025).

Delta Lake, developed by Databricks, provides transactional integrity and version control in cloud data storage systems. The central component of the architecture is a special log that records all changes in data and schemas, providing version control, rollback to previous states, and schema compliance checks (Delta Lake, 2025).

Each of these formats integrates with Apache Spark, a distributed computing platform for high-performance data processing. Since loading speed is a key performance indicator, comparing it under the same conditions allows us to identify the advantages and limitations of each format. Therefore, it is relevant to establish criteria for selecting the most appropriate Data Lakehouse architecture for building analytical data systems, considering the types and volumes of data, their loading speed, and disk space requirements with identical Spark settings and software/hardware conditions.

**Analysis of Recent Studies and Publications.** An experimental comparison of Apache Iceberg, Apache Hudi, and Delta Lake in terms of data loading, updating, and reading speeds was conducted in a study (Janssen, 2024). The authors presented their own LakehouseBench framework, which provides uniform testing conditions. The results illustrate that Delta Lake has better performance when loading data, Hudi when updating, and Iceberg provides high efficiency in complex analytical queries.

In a study (Jain, 2023), the authors evaluated the performance of Apache Hudi, Apache Iceberg, and Delta Lake on the Apache Spark platform in different scenarios: loading, updating, merging, and reading data. The results showed that Hudi performs better with frequent write and update operations, Iceberg provides the highest stability when reading large amounts of data, and Delta Lake demonstrated balanced performance in all modes.

The work (Begoli, 2021) is devoted to the application of Lakehouse architecture in biomedical systems with large amounts of data. Experiments with 50.8 million rows showed that using Parquet files instead of CSV improved query performance by approximately 50% and reduced storage volume by almost 10 times. The results confirm that Lakehouse systems not only reduce redundancy and costs but also provide scalable analytics for biomedical research.

A practical solution for streaming data from Apache Kafka to HDFS using Apache Spark was presented in (Drohobytskiy, 2024). The solution involves stream management, offset control, and the use of Parquet files. The study is an example of the effective organization of ingestion processes in environments similar to Lakehouse systems. The study (Fedorovych, 2024) also shows Spark's continuous and microbatch processing modes demonstrate fundamentally different latency and throughput characteristics, emphasizing the importance of choosing a data ingestion strategy consistent with the performance constraints of Lakehouse-oriented data pipelines.

Research on cloud deployments of Lakehouse systems using industry case studies was conducted in (Manchana, 2023). Their research showed that the implementation of a cloud Lakehouse solution improved scalability, reduced analytics execution from hours to minutes, and improved the quality of operational decisions.

A comparison of the efficiency of loading large amounts of data using Apache Spark, using the Java, Python, and Scala programming languages in the Apache Iceberg format table was performed in the article (Borodii, 2025). The results showed that Python is the fastest for small data sizes, while Scala demonstrates the best performance for large and complex ETL operations. The study illustrates the choice of programming language significantly affects the performance of Spark algorithms and should be considered when developing scalable data processing systems.

The previous studies show the feasibility of using Data Lakehouse architecture for different types of tasks, which can improve data processing performance. The research in this work focuses on analyzing loading time and disk space usage for different amounts of data. This is of practical importance for designing ETL processes in real analytical systems to choose the most suitable Lakehouse architecture.

**Purpose of the Article** is to define the most appropriate Data Lakehouse architecture based on the type and volume of data sources, data loading performance, and disk space required for analytical data systems with identical settings and software/hardware conditions. Based on the analysis, recommendations for using each of the studied Data Lakehouse systems will be formulated.

**Presentation of the Main Research Material.** This research compares the performance of the most common Data Lakehouse systems using Apache Spark as a tool for processing data arrays to select the most optimal architecture. Data Lakehouse systems differ in performance at the stage of primary loading of structured and semi-structured datasets. The study used versions Hudi 1.1.0, Iceberg 1.10.0, and Delta Lake 4.0.0, compiled under Scala 2.13, which ensures their compatibility with Apache Spark 4.0.1. The study was implemented in Python 3.11 using the PySpark library within the PyCharm environment. The experiment was conducted on a computer running Windows 10 Home (64-bit) with an AMD Ryzen 5 4600H processor and 16 GB of RAM. The JVM configuration provided 12288 MB of memory for the heap and concurrent execution of up to 12 tasks in local mode, where all computations are performed within a single physical node.

Several sources of structured and semi-structured data were used to form the input datasets. Structured data was obtained from the open-meteo web resource (Open-Meteo, 2025), which provides hourly air temperature indicators for 2022-2024 with reference to geographical coordinates. This dataset was downloaded into a 7 GB CSV file using API requests. Additionally, the study used a CSV file from the simplemaps.com web site (Simplemaps, 2025), which contains a list of cities and their coordinates, approximately 5 MB in size. Semi-structured data was obtained from the resource iqair.com (IQAir, 2025), that includes information on atmospheric air quality indicators depending on geographic coordinates. Access to the data was provided using the AirVisual API, and the results were obtained in JSON format.

All data was loaded into tables in each Data Lakehouse system using an ETL process to analyze the volume of the generated tables and evaluate the efficiency of disk space usage by each Lakehouse. Within the scope of this study, data extraction consisted of reading data from input files. During the transformation stage, the data and its structure were modified according to the requirements of the study using the Apache Spark data processing tool. The final stage, loading, involved transferring the prepared data to Data Lakehouse tables. The process of loading data into each of the Lakehouse systems consisted of the following stages: loading a list of cities, importing a file with hourly temperature readings according to geographical coordinates, obtaining data on atmospheric air quality, and combining three files by coordinates to load the generated dataset into a separate table.

For each of the Data Lakehouse systems under research, a separate database schema was created, which includes four tables for data loading. Each table corresponds to different aspects of data analytics. In particular, the weather_events table contains hourly temperature readings obtained from the open-meteo web resource. The world_cities table contains data on cities obtained from the simplemaps.com web site. The table is divided into partitions by country field in ISO2 format. The air_quality_level table contains data obtained from the AirVisual API, which includes nested struct objects with values for weather and pollution indicators. Additional fields lng, lat, ts, dt, and raw_json are generated during ETL process to simplify data storage and filtering. The data is partitioned by the dt field.

The weather_hourly_events entity contains data from three input files, combined by the lat and lng geographic coordinate fields. The table is partitioned by the attributes of year and month in yyyy_MM format, and country code in ISO2 format. The structure of the weather_hourly_events table is shown in Table 1.

Each Data Lakehouse system requires its own Spark configuration parameters to ensure correct execution of write operations and metadata management. The specific Spark configuration settings for each Lakehouse are listed in Table 2.

The other Spark configuration parameters are identical for each of the Data Lakehouse are listed in Table 3.

As part of the study, four separate experiments were conducted, each implementing an ETL process that involved the stages of data extraction from input files, its processing, and loading into the corresponding Data Lakehouse system using the Pyspark library. At the extraction stage, data was loaded from input files into DataFrame variables. At the transformation stage, operations are performed to reshape the data into a consistent format according to the data structure definition in Lakehouse. In specific, the to_date() date conversion function, the function of formatting temperature_yr_mnt field – date_format(), and the join() function for combining multiple datasets by the lat and lng coordinate fields were used. To optimize the combination of datasets, the broadcast() function was used, which allows small tables to be loaded into RAM.

Writing data to tables has been performed in overwrite mode using the partitionBy() method, which determines how the loaded data is organized in a file system. The additional save() parameter writes data directly to a Lakehouse table. For each Lakehouse system, writing data to tables has been implemented considering the characteristics of each system. For example, for Iceberg and Delta Lake, the standard modes write.format("iceberg") and write.format("delta") were used, while for Hudi, the bulk_insert mode and a COPY_ON_WRITE table format with the settings recordkey.field, precombine.field, and partitionpath.field were used to ensure correct identification of records and formation of the partition structure. The general structure of the data loading in Data Lakehouse systems is shown in Figure 1.

Table 1
**Structure of weather_hourly_events table**

| Field name | Data type | Is partitioned? |
|---|---|---|
| lat | double | No |
| lng | double | No |
| temperature_ts | timestamp | No |
| temperature | double | No |
| population | bigint | No |
| country | string | No |
| province | string | No |
| city | string | No |
| current | struct<pollution: struct<ts: timestamp, aqius: int, mainus: string, aqicn: int, maincn: string>, weather: struct<ts: timestamp, ic: string, hu: int, pr: int, tp: int, wd: int, ws: double, heatIndex: int> weather: STRUCT< ts: TIMESTAMP,ic: STRING, hu: INT,pr: INT, tp: INT,wd: INT, ws: DOUBLE, heatIndex: INT >> | No |
| temperature_dt | date | No |
| temperature_yr_mnt | string | Yes |
| country_iso2 | string | Yes |

Table 2
**Features of Spark configurations for each Data Lakehouse**

| Data Lakehouse | Configuration parameter | Description |
|---|---|---|
| Apache Iceberg | spark.sql.catalog.local = org.apache.iceberg.spark.SparkCatalog | Creates a local catalog to work with Apache Iceberg. |
| | spark.sql.catalog.local.type = hadoop | Specifies the catalog type (Hadoop means to create a storage based on the file system). |
| | org.apache.iceberg.spark.extensions.IcebergSparkSessionExtensions | Adds extensions to support Iceberg syntax in Spark (CREATE TABLE, MERGE INTO, etc.). |
| Apache Hudi | spark.sql.extensions = org.apache.spark.sql.hudi.HoodieSparkSessionExtension | Enables Hudi extensions in Spark to work with Copy-on-Write or Merge-on-Read table types. |
| | spark.sql.catalog.spark_catalog = org.apache.spark.sql.hudi.catalog.HoodieCatalog | Uses HoodieCatalog as the default Spark catalog. |
| Delta Lake | spark.sql.catalog.spark_catalog = org.apache.spark.sql.delta.catalog.DeltaCatalog | Sets DeltaCatalog as the default Spark catalog. |
| | spark.sql.extensions = io.delta.sql.DeltaSparkSessionExtension | Enables the Delta extension to support Delta Lake SQL commands such as MERGE, VACUUM, OPTIMIZE. |
| | spark.databricks.delta.retentionDurationCheck.enabled = false | Disables file retention checks during cleanup. |

**Results.** The research included four testing stages, varying in data volume, input data type (structured CSV and semi-structured JSON), ETL implementation complexity, and the resulting storage size of the Lakehouse tables. Table 4 shows the execution times of the ETL processes for loading each dataset into the Lakehouse tables using Apache Spark.

Although each table was loaded with the same input datasets, the resulting storage size differed across the Data Lakehouse systems. Table 5 shows the memory size for each Lakehouse table.

The results of the experimental comparison of Data Lakehouse systems in terms of processing time in seconds and memory capacity in megabytes for each table are shown in Figure 2.

The results of the experiment demonstrate that Delta Lake is the most efficient in data loading tasks regardless of the size of the dataset. In particular, the loading time for the largest dataset weather_hourly_events took 285.18 seconds, which is 26.6% faster than Apache Iceberg (388.55 seconds) and more than 2.5 times faster than Apache Hudi (737.03 seconds). In operations with smaller datasets, Delta Lake's advantage is less marked, but it remains, indicating a high level of optimization of the system for partitioned loading.

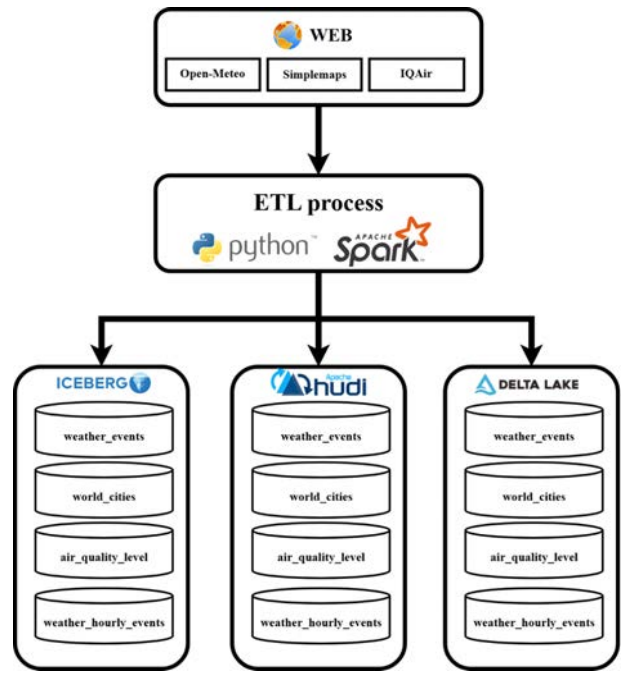

**Fig. 1. General structure diagram of data loading process in Data Lakehouse systems**

Apache Iceberg showed high performance, especially when working with smaller and medium-sized datasets. For air_quality_level and world_cities tables, loading times were 10.7 s and 6.16 s, respectively, which is only slightly different from Delta Lake's performance with 11.8% and 4.3%, accordingly. In the case of the large weather_events dataset (7 GB), Iceberg completed the load in 188.53 seconds, which is 18.2% slower

Table 3

**Spark configurations**

| Configuration parameter | Description |
| --- | --- |
| master = local[*] | Allows the Spark application to run in local mode using a multi-threaded approach. |
| spark.driver.memory = 12g | Sets 12 GB for the Spark driver process to manage tasks, metadata, and job coordination. |
| spark.sql.shuffle.partitions = 128 | 128 output partitions that Spark creates during data shuffling. |
| spark.sql.adaptive.enabled = true | Allows adaptive query execution. |
| spark.sql.adaptive.coalescePartitions.enabled = true | Automatically reduces the number of small partitions. |
| spark.sql.adaptive.skewJoin.enabled = true | Optimizes uneven connections when data distribution is uneven. |
| spark.serializer = org.apache.spark.serializer.KryoSerializer | Sets Kryo as the serializer type for data processing. |

Table 4

**Performance comparison of Data Lakehouse systems in Spark ETL**

| Table name | Apache Iceberg (sec) | Apache Hudi (sec) | Delta Lake (sec) |
| --- | --- | --- | --- |
| air_quality_level | 10.7 | 69.33 | **9.57** |
| world_cities | **6.16** | 10.01 | 6.44 |
| weather_events | 188.53 | 509.27 | **159.46** |
| weather_hourly_events | 388.55 | 737.03 | **285.18** |

Table 5
**Storage Size of Data Lakehouse Tables**

| Table name | Apache Iceberg | Apache Hudi | Delta Lake |
|---|---|---|---|
| air_quality_level | **17 MB** | 474 MB | 19.7 MB |
| world_cities | **1.94 MB** | 103 MB | 3.21 MB |
| weather_events | **248 MB** | 2.26 GB | 286 MB |
| weather_hourly_events | **270 MB** | 4.41 GB | 306 MB |

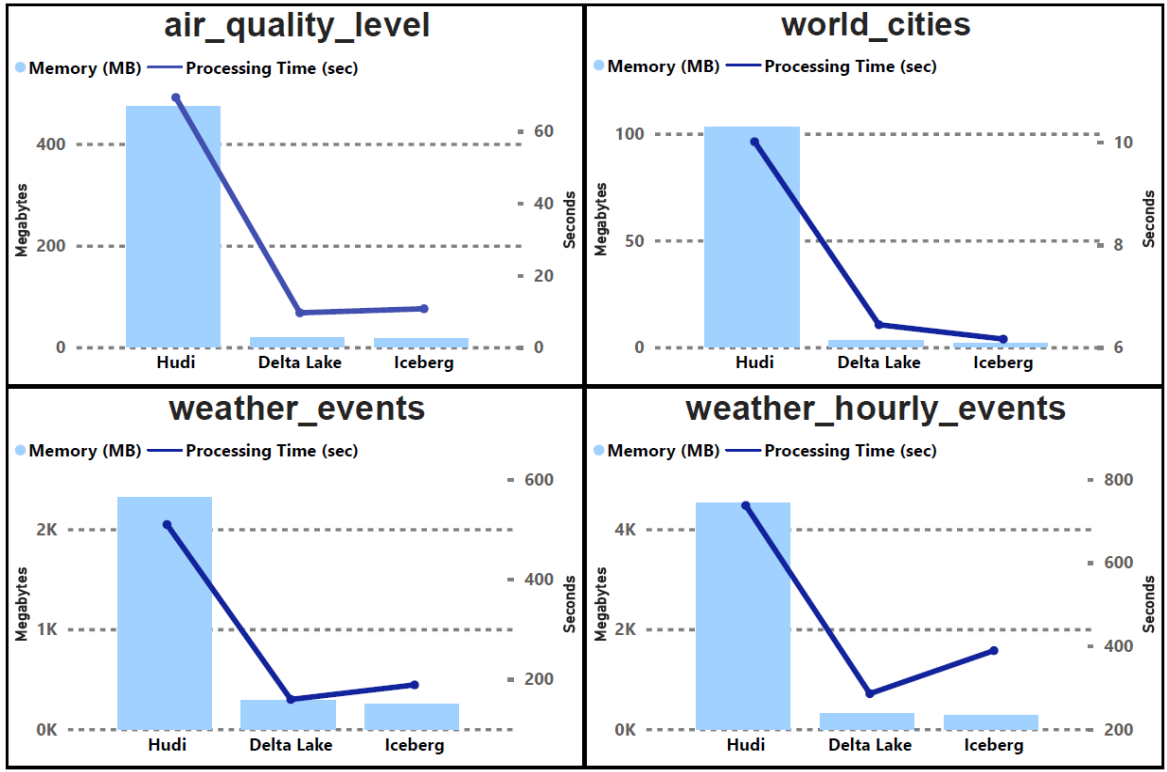

**Fig. 2. Comparative review of memory and performance for each Lakehouse**

than Delta Lake. At the same time, Apache Iceberg demonstrated the highest efficiency in terms of disk space usage. For example, after loading, the weather_hourly_events table took up only 270 MB in Iceberg, while in Delta Lake it took up 306 MB, and in Apache Hudi it took up 4.41 GB, which is more than 16 times as much. A similar trend was observed for other tables.

Apache Hudi showed the lowest efficiency among the three tested Lakehouse systems in terms of both data loading time and the size of the tables created. Even when working with smaller tables, such as air_quality_level (69.33 s) and world_cities (10.01 s), Hudi consistently lagged: 7.2 and 1.6 times slower than Delta Lake, respectively. Hudi's inefficiency in memory usage is particularly critical. In specific, the size of the weather_hourly_events table reached 4.41 GB, which is significantly larger than Iceberg (270 MB) and Delta Lake (306 MB).

**Conclusions.** This study compares the ETL performance of the three most widely used Data Lakehouse systems, such as Apache Iceberg, Apache Hudi, and Delta Lake including differences in data type and volume, loading speed, and resulting disk space usage under identical configurations. This is essential for analytical data system design.

The experimental results showed significant differences in system efficiency. Delta Lake demonstrated the fastest loading time in all research tasks, which is especially noticeable in case of large data volumes. Apache Iceberg showed stable performance and achieved results close to Delta Lake for small and medium data volumes. In addition, Iceberg provided the best disk optimization among the architectures studied. Apache Hudi, on the other hand, significantly lagged its competitors in terms of processing speed and table size, which was particularly evident in integrated scenarios.

The results indicate that the choice of a Lakehouse system should be based on the type of workload. Delta Lake is the most optimal solution for scenarios where the priority is fast and efficient loading of any data volumes. Apache Iceberg is best used in scenarios where stability and disk space savings are critical, especially when working with small and medium-sized datasets. Apache Hudi, on the other hand, is less suitable for high-volume data loading, but potentially effective in scenarios involving incremental updates and streaming processing.

**Prospects for further research.** In further research, the study will expand the experimental part to other aspects of data processing, such as updating, merging, deleting, and streaming data. Future research will involve the creation of a test environment based on Raspberry Pi, which will simulate the full data lifecycle: initial loading of historical arrays into a Lakehouse system, followed by periodic incremental updates on a daily basis. This will enable testing the performance of the systems in a context of limited hardware resources and scenarios closer to real-world operations.

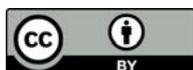